\documentclass[
  twocolumn,
  prb,
  showpacs,
  amsmath,
  amssymb,
  superscriptaddress,
  floatfix
]{revtex4-1}

\usepackage{bm}
\usepackage{graphicx}
\usepackage[bookmarks, colorlinks=true, plainpages = false,citecolor = blue, urlcolor = black, filecolor = blue]{hyperref}

\newcommand{\s}{\sum\limits}

\newcommand{\be}{\begin{equation}}
\newcommand{\e}{\end{equation}}
\newcommand{\beml}{\begin{subequations}}
\newcommand{\eml}{\end{subequations}}
\newcommand{\beq}{\begin{eqnarray}}
\newcommand{\eq}{\end{eqnarray}}
\newcommand{\bal}{\begin{align}}
\newcommand{\eal}{\end{align}}
\newcommand{\ba}{\begin{array}}
\newcommand{\ea}{\end{array}}
\newcommand{\bpm}{\begin{pmatrix}} 
\newcommand{\epm}{\end{pmatrix}} 
\newcommand{\bc}{\begin{cases}} 
\newcommand{\ec}{\end{cases}} 
\newcommand{\lt}{\left}
\newcommand{\rt}{\right}

\newcommand{\bb}{\mathbf}
\newcommand{\bs}{\boldsymbol}

\DeclareMathOperator{\tr}{Tr}

\DeclareMathOperator{\im}{Im}

\begin{document}

\title{Electron-hole asymmetry in two-terminal graphene devices}
\author{W.-R. Hannes}
\affiliation{SUPA, Department of Physics, Heriot-Watt University, Edinburgh EH14 4AS, UK}
\author{M. Jonson}
\affiliation{SUPA, Department of Physics, Heriot-Watt University, Edinburgh EH14 4AS, UK}
\affiliation{Department of Physics, University of Gothenburg, SE-412 96 G\"{o}teborg, Sweden}
\affiliation{Department of Physics, Division of Quantum Phases \& Devices, Konkuk University, Seoul 143-701, Korea}
\author{M. Titov}
\affiliation{SUPA, Department of Physics, Heriot-Watt University, Edinburgh EH14 4AS, UK}
\date{10 May 2011}
\begin{abstract}
A theoretical model is proposed to describe asymmetric gate-voltage dependence of conductance and noise in two-terminal ballistic graphene devices. The model is analyzed independently within the self-consistent Hartree and Thomas-Fermi approximations. Our results justify the prominent role of metal contacts in recent experiments with suspended graphene flakes. The contact-induced electrostatic potentials in graphene demonstrate a power-law decay with the exponent varying from $-1$ to $-0.5$. Within our model we explain electron-hole asymmetry and strong Fabri-Perot oscillations of the conductance and noise at positive doping, which were observed in many experiments with submicrometer samples. Limitations of the Thomas-Fermi approximation in a vicinity of the Dirac point are discussed. \end{abstract}
\maketitle

\section{Introduction}

The recent observation of the fractional quantum Hall effect in graphene\cite{Du:2009} calls for better understanding of two-terminal transport measurements in suspended graphene devices. The two-terminal geometry has also been used to reach the ballistic regime of transport with submicrometer flakes on SiO$_2$ substrate.\cite{Heersche:2007,Danneau:2008} These and other experiments\cite{Du:2008,Du:2008:NN} share such common features as the electron-hole asymmetry in the gate-voltage dependence of the conductance and the prominent conductance oscillations at positive doping that are beyond a simplistic theory of ballistic transport.\cite{Tworzydlo:2006,Titov:2006} In this paper we attribute these experimental observations to the effect of charge transfer in a vicinity of metal electrodes and to a weak screening in graphene at low doping.

The influence of metal contacts on electron transport in nearly ballistic graphene has been investigated both experimentally\cite{Lee:2008,Huard:2008,Blake:2009,Mueller:2009,Russo:2010} and theoretically.\cite{Khomyakov:2009,Khomyakov:2010,Vanin:2010,Zhang:2008,Datta:2009,BarrazaLopez:2010,Nouchi:2010} An agreement between theory and experiment has, however, been limited due to both the complexity of experimental setups and a lack of important ingredients in the considered theoretical models. In this study we suggest a minimal model, which leads to a very good description of experimental data by Xu Du \textit{et~al.}\cite{Du:2008:NN} on charge transport in suspended graphene. The gate-voltage dependence of transport properties predicted by the model also agree qualitatively with those observed by Heersche \textit{et~al.}\cite{Heersche:2007} for graphene on SiO$_2$ substrate.


The electrostatic potential landscape in graphene with metal contacts has been imaged by means of scanning photocurrent microscopy\cite{Lee:2008,Mueller:2009} to demonstrate the long-range decay of the contact-induced potential (about 1 $\mu$m). These experiments and the theory\cite{Nouchi:2010} suggest that the charge density in graphene covered by a metal is pinned, i.e. not affected by the gate voltage. 

The pinning is much weaker if graphene sheet is oxidized and the metal-graphene contact is resistive. In this case the second minimum in the gate voltage dependence of conductance can emerge which corresponds to the charge density minimum in the metal-covered graphene. In Ref.~\onlinecite{Nouchi:2010} this mechanism is claimed to be responsible for the electron-hole asymmetry observed in Refs.~\onlinecite{Nouchi:2008,Du:2008}. We argue, however, that the asymmetry seen in Refs.~\onlinecite{Du:2008,Du:2008:NN,Du:2009} is entirely due to the contact-induced potential in the free-standing graphene. We also regard the observed saturation of the conductance for large negative gate voltages as the signature of the charge density pinning in the metal-covered graphene. The same mechanism of the electron-hole asymmetry has been put forward recently in Ref.~\onlinecite{Nouchi:2011}.

The adsorption of graphene on metal substrates has been studied by means of the density functional theory (DFT) in Refs.~\onlinecite{Khomyakov:2009,Vanin:2010}. These studies suggest that the main effect of metal deposition can be modeled by a shift of the Fermi level, i.e. by the chemical doping. This shift can be of any sign. Its precise value, $\mu_l$, is comparatively large due to the small density of states of intrinsic graphene, and depends on the workfunctions of graphene and the metal as well as on the direct chemical interaction. The most recent analysis\cite{Vanin:2010} predicts, e.g. $\mu_l=-0.51$\,eV for Al, $-0.40$\,eV for Ag, $-0.43$\,eV for Cu, and $0.21$\,eV for Au. 

In this paper we refer to the metal-covered graphene as the lead, and to the free-standing graphene as the sample. The scalar step-function potential in the Dirac equation provides the simplest model of the sample-lead interface.\cite{Tworzydlo:2006,Titov:2006} Remarkably, the exact analytical solution can also be found if the interface is modeled by an exponentially decaying potential.\cite{Cayssol:2009} These models, however, are not self-consistent and cannot describe the large electron-hole asymmetry observed in experiments.

Microscopic DFT-modeling of the sample-lead interface has been undertaken in Ref.~\onlinecite{BarrazaLopez:2010} to investigate the contact-induced potential in short ($L<14$nm) graphene samples, and its effect on the two-terminal conductance. In this study a gate electrode is absent and the conductance is studied as a function of Fermi energy using electrostatic potential for undoped graphene. The reported electron-hole asymmetry is, therefore, of a different type than the one occurring with pinned charge density in the leads. It is also hard to scale the results of Ref.~\onlinecite{BarrazaLopez:2010} to realistic system sizes ($L>100$\,nm). 

Below we propose an effective model, which takes into account both the charge density pinning in the leads and the self-consistent contact potential in the sample. In formulating the model we focus on the two-terminal geometry and neglect the resistance of the sample-lead interfaces. We employ the effective Dirac Hamiltonian and take into account the large density of states in a metal. The gate electrode is incorporated into the model, while the system size poses no serious restriction. The presence or absence of a substrate is taken into account by an appropriate choice of the permittivity constant. We calculate the self-consistent potential in graphene using two different approximations, namely the self-consistent Hartree (SCH) and the Thomas-Fermi (TF) approximation. From the Kubo formula we, then, calculate the two-terminal conductance and noise as a function of gate voltage. 

We find that the different approximations result in a qualitatively similar form of the effective potential. The potential penetrates deeply into the sample and reveals a power-law decay with the exponent varying from $-1$ to $-0.5$ depending on the electron concentration. This behavior agrees with the TF analysis of Ref.~\onlinecite{Khomyakov:2010}. We demonstrate that the slow decay of the potential is responsible for the asymmetric gate-voltage dependence of conductance and noise, increasing in intensity with the lead doping $|\mu_l|$. For small positive gate voltages the Dirac point crosses the chemical potential twice inside the sample so that two n-p interfaces are formed. The slow potential decay ensures that the transmission through the interfaces is strongly selective with respect to the momentum direction.\cite{Cheianov:2006} The electron scattering at the n-p interfaces is the reason for the enhanced Fabry-P{\'e}rot oscillations of the conductance as a function of electron concentration at moderate positive doping. We note that the oscillations transform to the pronounced resonances in the ribbon geometry $W< L$.\cite{Darancet:2009} When describing gate-voltage asymmetries we refer to negatively doped leads, $\mu_l<0$. For positively doped leads the polarity of the effect is reversed. One can further argue that Fabry-P{\'e}rot oscillations are easily distorted by disorder. The presence and shape of the Fabry-P{\'e}rot oscillations can, therefore, be used as a direct indicator of the sample quality. 

\section{Charge and potential profiles}

The charge carriers in the graphene sheet (defining the $xy$-plane) are described by the Dirac Hamiltonian 
\be
\label{H}
H = H_0 + V(x),\qquad H_0=-i\hbar v \bs{\sigma\cdot \nabla},
\e
where $\bs{\sigma}$ is a vector of Pauli matrices and $V(x)$ is the spatially dependent effective potential. In Eq.~\eqref{H} we neglect any effects arising from the finite band width and assume translational invariance in $y$.

\begin{figure}[t!]
\includegraphics[width=0.85\columnwidth]{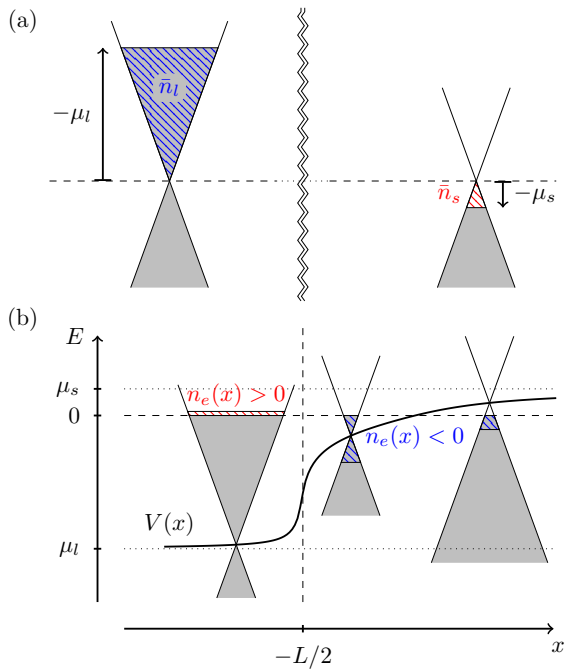} 
\caption{(Color online) Schematic illustration of the doping in the left lead (metal-covered graphene) and the sample (ballistic graphene)  if they (a) are not and (b) are in electrical contact with each other. The indicated excess charge density $n_{e}(x)$ in (b) (hashed regions) corresponds to the semiclassical (local) approximation. Quantum mechanical (non-local) corrections to $n_{e}(x)$ exist in the gray-shaded regions (Friedel oscillations) and in regions where the transverse momentum is outside the Dirac cone (evanescent modes).}
\label{fig:contactpot}
\end{figure}

Doping of the graphene sheet is caused by the contact with the metal electrodes and by applying a voltage to the global back gate. We model the lead by graphene with the chemical potential $\mu_l$ measured from the Dirac point as shown in Fig.~\ref{fig:contactpot}(a).

We use the externally compensated charge density $\bar n(x)$ in graphene as an input parameter for a self-consistent calculation. We focus on a two-terminal setup for which the metal-covered and free-standing parts of the graphene sheet correspond to  $|x|>L/2$ (leads) and  $|x|\leq L/2$ (sample), respectively. Accordingly, the compensated charge density has the spatial dependence
\be
\bar n(x)=\bc \bar n_s,& |x|<L/2\\ \bar n_l,& |x|>L/2 \ec,
\e
where $\bar n_s$ is proportional to the gate voltage, since the distance $d_\mathrm{g}$  to the gate electrode is typically such that the quantum corrections to capacitance can be neglected. We assume that the charge density in the lead, $\bar n_l$, is pinned,\cite{Nouchi:2010} i.e. not influenced by the gate-voltage. The screening inside the gate electrode can be neglected for $L\lesssim \kappa_\mathrm{s}\, d_\mathrm{g}$, with $\kappa_\mathrm{s}$ the relative permittivity of the gate dielectric. Finally, the inter-valley coupling at the metal-graphene interface is also disregarded.

The relation between the charge densities $\bar n_s$ and $\bar n_l$ and their respective chemical potentials $-\mu_s$ and $-\mu_l$ (cf. Fig.~\ref{fig:contactpot}(a)) is given by
\be\label{nmu}
\bar n_{s,l} = \int dE\ \rho_0(E)\, \big\{ (1\!-\!\theta_E) (1\!-\! f_{E+\mu_{s,l}}) - \theta_E f_{E+\mu_{s,l}} \big\},
\e
where $f_E$ is the Fermi distribution function, $\theta_E$ is the Heaviside step function, and 
\begin{equation}\label{DOS}
\rho_0(E)= \frac{2 |E|}{\pi\hbar^2 v^2}
\end{equation}
is the density of states (DOS) in infinitely extended ballistic graphene described by Hamiltonian~\eqref{H} including the spin and valley degeneracy. At zero-temperature Eq.~\eqref{nmu} becomes
\begin{equation}\label{nmu}
n_{s,l} =  \frac{\mu_{s,l}|\mu_{s,l}|}{\pi \hbar^2 v^2}.
\end{equation}  

Screening in the leads is strongly enhanced by the metal electrodes deposited on top of the graphene sheet. This effect, which ensures the charge-density pinning, is taken into account by means of an additional DOS  $\rho_\mathrm{m} \gg \rho_0(\mu_{l})$, which is assumed to be energy independent. The transport properties do not depend on the value of $\rho_\mathrm{m}$ under these conditions. 

The density profile $\bar n(x)$ corresponds to a charge-neutral setup with disconnected leads. Bringing the sample into electric contact with the leads causes the charge redistribution and the band bending shown in Fig.~\ref{fig:contactpot}(b). The excess charge density $n_{e}(x)$ in the entire device is defined as
\be\label{n_e}
n_{e}(x)= n(x)+ n_\mathrm{m}(x) -\bar n(x),
\e
where $n(x)$ is the total charge density in the graphene sheet relative to intrinsic graphene, and 
\be
n_\mathrm{m}(x) = \lt(V(x) - \mu_l \rt)\,\rho_\mathrm{m}\,\theta_{|x|-L/2}
\e
is the additional charge density induced in the metal.

The potential profile $V(x)$ and the total charge density $n(x)$ have to be determined self-consistently. In this study we ignore the exchange interaction and restrict ourselves to the scalar Hartree potential
\be\label{V}
V(x)=\mu_l+V_\mathrm{H}(x),
\e
where
\be
\label{int}
V_H(x) =  \alpha \hbar v \!\! \int_{-\infty}^{\infty} \!\!\!\! dx'\, n_{e}(x')\!\!
\int_{-W/2}^{W/2}  \!\!\!dy\ \frac{e^{-r(x-x',y)/a_0}}{r(x-x',y)}.
\e
The coupling constant is given by $\alpha=\alpha_0/\kappa_\mathrm{bg}$, where  $\alpha_0=e^2/4\pi\epsilon_0\hbar v \approx (c/v)\times(1/137) \approx 2.2$ is the fine structure constant in ballistic graphene, and $\kappa_\mathrm{bg}$ is the effective background dielectric constant in the graphene plane ($\kappa_\mathrm{bg}\!=\! 1$ for suspended graphene and $\kappa_\mathrm{bg}\!\approx\!2.0$ for graphene on  SiO$_2$ with the other side exposed to air/vacuum). The distance $r(x,y)$ depends on the ``geometry'' of the graphene sheet. In a planar geometry, $r=\sqrt{x^2+y^2}$, one has to regularize the interaction term by choosing a finite screening length $a_0$, which requires that the extension of the leads in $x$ is much smaller than $a_0$ in order to fulfill the charge neutrality condition. Alternatively one can let $a_0\to \infty$ by considering a graphene sheet in the form of a cylinder with circumference $W\gg L$. This choice yields the same results for $V(x)$ with a computational advantage.   

The self-consistent scheme is closed by relating the total charge density $n(x)$ to the potential profile $V(x)$. 
The quantum-mechanical expression for $n(x)$ in the self-consistent Hartree (SCH) approximation reads
\be\label{SCH}
n_\mathrm{SCH}(x) = \int dE\, f_{E-V(x)}\, \rho_0(E-V(x)) - f_{E} \rho(x;E).
\e
The local density of states (LDOS), $\rho(x;E)= - (4/\pi)\im \tr G^R(\bb{r},\bb{r};E)$, is calculated by solving the equation for the retarded Green's function, $G^R$,
\begin{equation}
\label{green}
(E+i\eta-H)\ G^R(\bb{r},\bb{r}';E)=\delta(\bb{r}-\bb{r}'),
\end{equation}
where $\eta$ is a small positive parameter. We note that the dependence of the LDOS on the effective potential is non-local. Taking advantage of the translational invariance in $y$ we introduce the Fourier transform
\begin{equation}
G^R(\bb{r},\bb{r}';E)=\frac{1}{W}\s_q \exp[{iq\,(y-y')}]\ G_{q}^R(x,x';E),
\end{equation}
where the summation runs over the discrete values of conserved transverse momentum, $q_n=2\pi n/W$, with $n$ integer (periodic boundary conditions in $y$ direction).

The boundary conditions in $x$ for  the Green's function $G_q^R(x,x';E)$ in the channel representation are obtained from the exact analytical solution for $|x|>\xi$ and $|x'|<\xi$, where $\xi>L/2$ is such that $V(|\xi|)\simeq \mu_l$.
Selecting the decaying solution ($G_q^R(x,x';E)\to 0$ for $x\to\pm\infty$) we take the limit $\eta\to 0$. Still, since the spectrum is partially discrete (some modes are confined to the sample), it is necessary to keep $\eta$ finite in order to maintain computational stability of the scheme. We consider the limit $\hbar v/W \ll \eta \ll \hbar v /L$ such that coherence is preserved on the scale $\hbar v/\eta \gg L$. 

The numerical computation of the Green's function $G_q^R(x,x;E)$ for $|x|<\xi$ is demanding since a large energy range has to be considered. For $\mu_l<\mu_s$ the lower energy bound at which the integrand in Eq.~\eqref{SCH} becomes negligible is roughly $\mu_l - 2\times (\mu_s - \mu_l)$. To ensure the charge conservation one has to keep the parameters $W$ and $\eta$ in Eq.~\eqref{SCH} identical for both $\rho(x;E)$ and $\rho_0(E)$, so that  
\begin{equation} \label{rho0}
\rho_0(E) = \frac{4}{\hbar v \pi W} \s_q \im\lt( \frac{E+i\eta}{\sqrt{-(E+i\eta)^2+(\hbar v\, q)^2}}\rt),
\end{equation}
instead of Eq.~(\ref{DOS}). The expression (\ref{rho0}) converges to Eq.~(\ref{DOS}) in the limit $W\to \infty$, $\eta\to 0$.


The relation between charge density and effective potential is local in the TF approximation, which dramatically improves the computational efficiency. Replacing $\rho(x;E)$ in Eq.~\eqref{SCH} by $\rho_0(E-V(x))$, setting the temperature to zero, and using the ideal DOS \eqref{DOS} leads to the semiclassical expression for $n(x)$,
\begin{equation}\label{TF}
n_\mathrm{TF}(x) = \frac{V(x) |V(x)|}{\pi \hbar^2 v^2},
\end{equation}  
which we regard as the charge density in the TF approximation.


The set of Eqs.~\eqref{V},\eqref{n_e}, along with Eq.~\eqref{SCH} for SCH and Eq.~\eqref{TF} for TF approximations, is solved by means of an iterative algorithm. It is run until $V(x)$ reaches a self-consistency with accuracy of 10$^{-3}$ relative to $|\mu_l|$. For strong interactions ($\alpha\sim 1$) and large potential steps the procedure requires the use of strong damping in each iteration to ensure convergence (the weight of a new iteration is roughly of the order of $10^{-3}$).

\begin{figure}[t!]
\includegraphics[width=0.85\columnwidth]{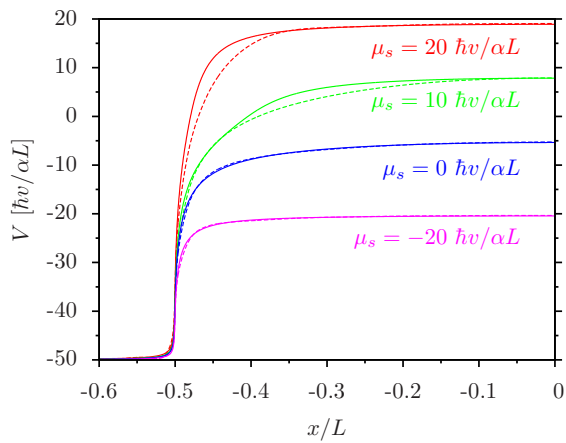} 
\caption{(Color online) Contact potentials, $V(x)$, (symmetric with respect to $x=0$) are calculated in the TF approximation (solid lines) and SCH approximation (dashed lines, $\alpha=1$) for different values of the sample doping, $\mu_{s}$, and fixed $\mu_{l}=-50\,\hbar v/\alpha L$ ($\alpha\rho_\mathrm{m}\hbar v L=100$, $W/L=200$). 
}
\label{fig:mu_muL-50}
\end{figure} 

Potential profiles $V(x)$ for different values of $\mu_s$ but fixed $\mu_l$, calculated at zero temperature, are shown in Fig.~\ref{fig:mu_muL-50}. The dimensionless potential $\alpha V(x) L/\hbar v$ in the TF approximation depends on the two parameters  $\alpha \mu_s L/\hbar v$ and $\alpha \mu_l L/\hbar v$. Such scaling is only approximate for the SCH potential. We also note that the TF potential transforms exactly as $V(x)\to -V(x)$ under the global transformation $\bar n(x)\to-\bar n(x)$, which is not the case for the SCH potential. 

Our results apply to large system sizes which are not accessible by the DFT models. For instance, for Al contacts ($\mu_l=-0.51$\,eV) and $L=250$\,nm one finds $\mu_l L/\hbar v \approx -200$. The sample width is of minor influence as far as $W > L$. Although the parameter $\rho_\mathrm{m}$ determines the potential decay in the leads (but not the charge profile), it has no significant effect on transport properties as long as $|\mu_{s}|<|\mu_{l}|$.

\begin{figure}[t!]
\includegraphics[width=0.98\columnwidth]{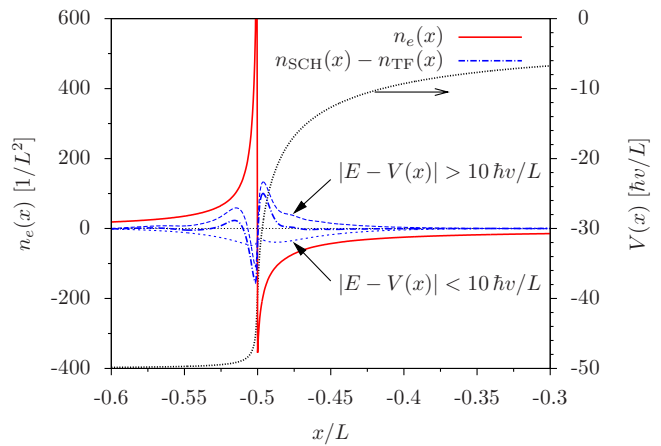} 
\caption{(Color online) The dotted curve (right axis) shows the self-consistent TF potential, $V(x)$, for $\mu_{s}=0$ and  $\mu_{l}=-50\,\hbar v/L$ ($\alpha=1$, $\rho_\mathrm{m}\hbar v L=100$, $W/L=200$). The corresponding TF excess charge density $n_e(x)$ is plotted with a solid line. The dash-dotted curve is the quantum correction to the charge density, obtained from the SCH method. This correction is split into contributions from energies close and further away from the Dirac point as shown by dashed lines.
}
\label{fig:x-n}
\end{figure} 

The validity of the TF approximation is governed by the semi-classical criterion $|d\lambda(x)/dx|/2\pi \ll 1$, where $\lambda(x)$ is the de Broglie wavelength. This is equivalent to the condition $|dV(x)/dx| \ll V^2(x) / \hbar v$,\cite{Khomyakov:2010} which is violated for the upper curve in Fig.~\ref{fig:mu_muL-50} ($\alpha\mu_s L/\hbar v =20$). The notable difference between the TF and SCH potential in this case indicates the importance of non-local quantum effects: evanescent modes and Friedel oscillations. Evanescent modes strongly increase the charge density in a vicinity of the sample-lead interface for energies close to the Dirac point.\cite{Titov:2010} This effect, however, is partially compensated in the considered geometry by the Friedel oscillations. The latter suppress the LDOS near the interface for energies far from the Dirac point. To illustrate the compensation we plot in Fig.~\ref{fig:x-n} the quantum correction $n_\mathrm{SCH}(x)-n_\mathrm{TF}(x)$ for energies $|E-V(x)|<10\,\hbar v/L$ (dominated by the evanescent modes) and  $|E-V(x)|>10\,\hbar v/L$ (dominated by the Friedel oscillations). The partial compensation of the non-local quantum corrections makes the TF approximation reliable even outside its applicability range. Still the mentioned discrepancy between the TF and SCH potentials shows the limitations of local density approximations such as TF in positively doped samples.\cite{comment} 

To make a direct comparison of our self-consistent calculation to the TF analysis of Ref.~\onlinecite{Khomyakov:2010}, we calculate the TF potential for a single sample-lead interface at $x=0$ ($\bar n(x)=\bar n_{l}$ for $x< 0$ and $\bar n(x)=\bar n_{s}$ for $x>0$). We find that the TF potential decays for $x>0$ as $x^{-p}$. The exponent $p$ is given by $p=1/2$ for $|V(x)/\mu_l| \ll 1$ and $p=1$ otherwise (which means that $p$ can vary with position). A similar behavior is found in Ref.~\onlinecite{Khomyakov:2010} when the doping inside the sample, $\mu_s$, is due to the charged impurities. Our results for gated graphene are different since no charge density pinning in the leads is assumed in Ref.~\onlinecite{Khomyakov:2010}.

\begin{figure}[t!]
\includegraphics[width=0.85\linewidth]{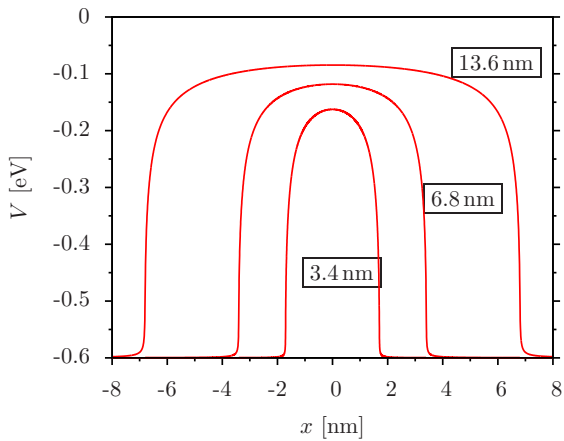} 
\caption{(Color online) Potential profiles $V(x)$ for $\mu_{l}=-0.6$\,eV, $\mu_{s}=0$, $\alpha=2.2$, and three different sample lengths $L =$ 3.4, 6.8, and 13.6\,nm chosen as in Ref.~\onlinecite{BarrazaLopez:2010}.
}
\label{fig:V_L-varied}
\end{figure} 

To compare our results to the full DFT treatment of Ref.~\onlinecite{BarrazaLopez:2010} we plot in Fig.~\ref{fig:V_L-varied} the self-consistent TF potentials calculated from our model for the same set of parameters $\mu_l=-0.6$\,eV, $\mu_s=0$, $\alpha=2.2$, and $L=$ 3.4, 6.8, and 13.6\,nm. Despite the simplicity of our model (notably, the absence of exchange interactions) the potentials agree well with those calculated in Ref.~\onlinecite{BarrazaLopez:2010}. 

\section{Transport properties}

For a given potential profile one may calculate the zero-temperature conductance from the Landauer formula $G = (4e^2/h) \sum_q T_q$, where the sum extends over the transverse momenta $q=2\pi n/W$. The shot noise is quantified by the Fano factor $F = \sum_q T_q (1-T_q) / \sum_q T_q$. 

The transmission probability $T_q=T_q(E=0)$ for a given channel at the Fermi energy is related to the Green's function by the Kubo formula\cite{Ostrovsky:2007}
\begin{equation}
T_q(E) = (\hbar v)^2 \tr\lt[ \sigma_x G_q^R(\xi,-\xi;E) \sigma_x G_q^{R\dagger}(\xi,-\xi;E) \rt],
\end{equation}
where $\xi$ is an optional cross-section. We choose $\xi$ such that $V(x)\simeq \mu_l$ for $|x|>\xi$.
 
\begin{figure}[t!]
\includegraphics[width=0.98\linewidth]{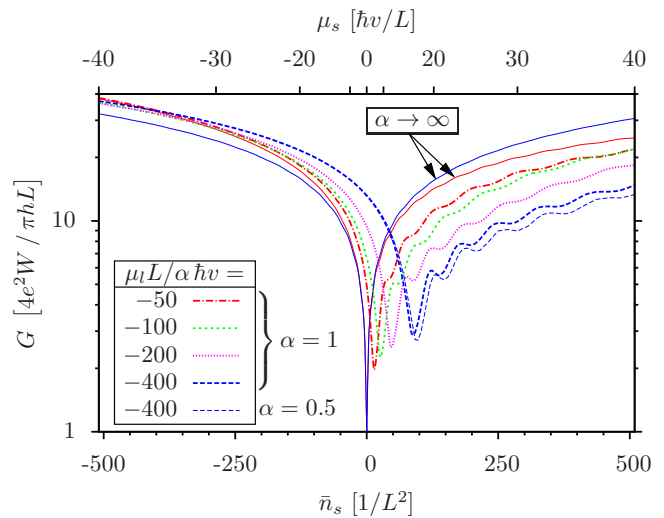} 
\caption{(Color online) Zero-temperature conductance as a function of external doping $\bar n_{s}$. The TF potential profiles are used. The thin solid lines marked with "$\alpha\to\infty$" correspond to step-like potentials, $V(x)=\mu_{l}$ for $|x|> L/2$ and $V(x)=\mu_{s}$ for $|x|<L/2$, with $\mu_l L / \hbar v$ given by -50 and -400 for the red (light) and blue (dark) line, respectively.}
\label{fig:nS-G}
\end{figure}

\begin{figure}[t!]
\includegraphics[width=0.98\linewidth]{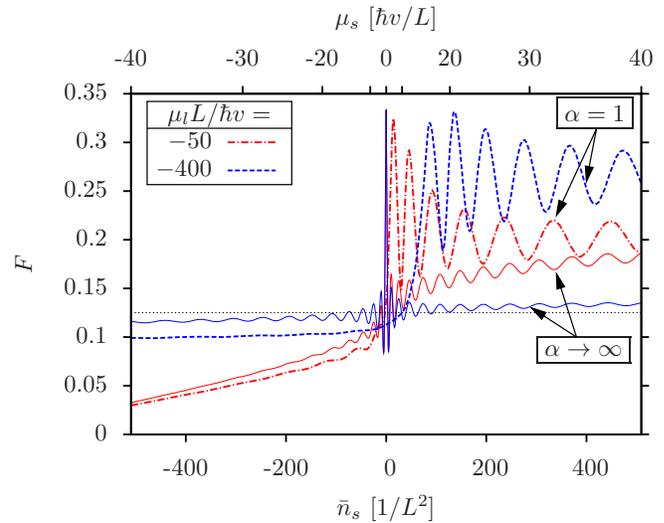} 
\caption{(Color online) Zero-temperature Fano factor as a function of external doping $\bar n_{s}$. 
As in Fig.~\ref{fig:nS-G}, the TF potential profiles are used (for the thick lines) and the thin solid lines correspond to the corresponding step-like potentials. The horizontal dotted line indicates the asymptotics $F=1/8$ for a step-like potential with $\mu_{l}\to\pm\infty$.\cite{Tworzydlo:2006}}
\label{fig:nS-F}
\end{figure}

The conductance in the TF approximation is shown in Fig.~\ref{fig:nS-G} as a function of the charge density in the sample, $\bar n_{s}$, for different values of $\mu_l$ (using the SCH approximation gives very similar results). We note that the conductance scales approximately with the parameter $\mu_l L/\alpha \hbar v$ for $|\mu_l| L/\alpha \hbar v\gg 1$. A similar plot of the Fano factor is given in Fig.~\ref{fig:nS-F}.

The most evident consequence of the charge transfer between the sample and the leads is an electron-hole asymmetry in the dependence of the conductance and noise on the charge density $\bar n_{s}$ (or the gate voltage). To illustrate the key role of the slow potential decay we plot in Figs.~\ref{fig:nS-G},\ref{fig:nS-F}, for comparison, the results obtained from the step-function model ($V(x)=\mu_{s}$ for $|x|< L/2$ and $V(x)=\mu_{l}$ for $|x|> L/2$) with thin solid lines. The step-function model also leads to an asymmetry, which is, however, negligible in the experimentally relevant regime $|\mu_l| L/\hbar v  \gg 1$, $|\mu_s| \ll |\mu_l|$.  

\begin{figure}[t]
\includegraphics[width=0.99\linewidth]{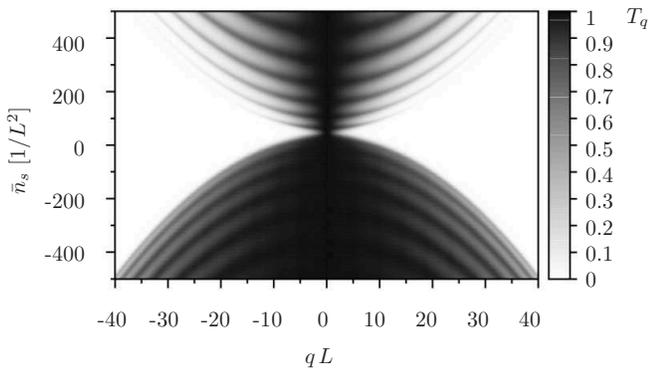} 
\caption{Zero-temperature transmission probability $T_q$ as a function of transverse momentum $q$ and doping $\bar n_{s}$ ($\alpha=1$, $\mu_{l} L/\hbar v=-200$).}
\label{fig:q-ns-T}
\end{figure}

\begin{figure}[t!]
\includegraphics[width=0.86\linewidth]{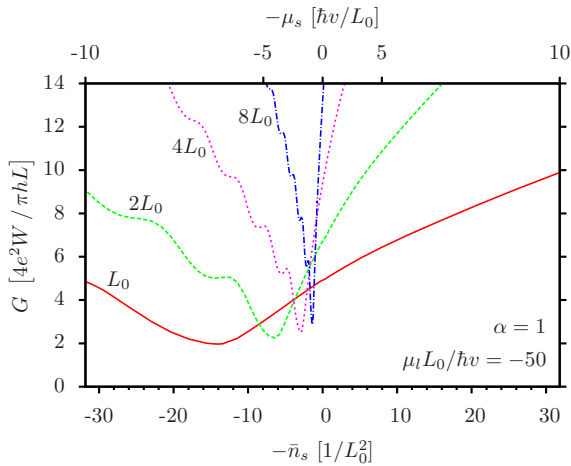} 
\caption{(Color online) Same conductance data as in Fig.~\ref{fig:nS-G} plotted with fixed $\mu_l$ and varying $L$, which is indicated  in multiples of the sample length $L=L_0$ for the solid curve. Note that the horizontal axis is inverted.}
\label{fig:nS-G_L-varied}
\end{figure}

We find that the conductance is enhanced for negative doping since $|V(x)|>|\mu_{s}|$ in the entire sample and suppressed for sufficiently large positive doping since $|V(x)|<|\mu_{s}|$ in this case. The minimal conductance is gradually increasing for large $\mu_l  L$, while the position of the conductivity minimum is shifted towards the positive doping. 

Another consequence of the slow potential decay is the specific form of the Fabry-P{\'e}rot oscillations, which are only visible on one side of the conductance minimum (in the Fano factor one can also observe weak oscillations for negative doping). As demonstrated by thin solid lines in Fig.~\ref{fig:nS-F}, a small asymmetry already exists in the step-function model with finite doping of the leads. However, the oscillation amplitude is strongly enhanced (reduced) for positive (negative) doping as a consequence of the slow potential decay. The enhancement is strongest when the position of the Dirac point, $V(x)$, coincides with the chemical potential in a spatially extended region, which (as illustrated in Fig.~\ref{fig:q-ns-T}) leads to the selective transmission of the carries with small transverse momenta at the n-p interfaces.\cite{Cheianov:2006} 

In a particular set of experiments it is the sample length, $L$, rather than the lead doping, that is varied. In Fig.~\ref{fig:nS-G_L-varied} we rearrange the data to show the dependence of conductance on the sample doping for different sample lengths. As expected the contact-induced electron-hole asymmetry is more pronounced for shorter samples. The shift of the conductance minimum qualitatively agrees with recent measurements reported in Ref.~\onlinecite{Nouchi:2011}.

\section{Conclusions}

In conclusion we have modeled two-terminal electron transport through ballistic graphene samples by taking into account the effects of charge transfer at the metal-contact/graphene interfaces. Our analysis explains the electron-hole asymmetry in the conductance and the Fabry-P{\'e}rot conductance oscillations at positive doping, which have been observed in many experiments.\cite{Heersche:2007,Du:2008,Du:2008:NN,Du:2009} These phenomena are most clearly resolved in Ref.~\onlinecite{Du:2008:NN} (Fig.\,2 b,c) for two different sample lengths. The period $\delta\mu_{s}$ of the oscillations, found by transforming the positions of the peaks into $\mu_{s}$, corresponds to $\delta\mu_{s} L/\hbar v \approx \pi$, which unambiguously confirms that the oscillations are of the Fabry-P{\'e}rot  type. The amplitude of the oscillations in this experiment is somewhat stronger than in Fig.~\ref{fig:nS-G}, which is likely due to a modification of graphene by the metal leads that has resulted in an increased reflectivity of the sample-lead interfaces.

We are grateful to I.\,V.\,Gornyi,  P.\,M.\,Ostrovsky, H.\,Schomerus, and J.\,Cayssol for discussions. WRH acknowledges funding by the Scottish Universities Physics Alliance. MJ acknowledges partial support from the Swedish VR and the Korean WCU program funded by MEST/NFR (Grant No. R31-2008-000-10057-0). MT acknowledges the hospitality of Karlsruhe Institute of Technology where part of the work was done.


\begin{thebibliography}{99}
\bibitem{comment}
Strictly speaking, there exists a purely technical reason for the difference between TF and SCH potentials since the latter is calculated for a finite broadening $\eta$. This effect is, however, negligible in Fig.~\ref{fig:mu_muL-50}.
\end{thebibliography}
\end{document}